\documentclass[a4paper]{article}

\usepackage{INTERSPEECH2021}
\usepackage{booktabs}
\usepackage{tabularx}
\usepackage{url}

\title{iMiGUE-Speech: A Spontaneous Speech Dataset for Affective Analysis} 
\name{Sofoklis Kakouros, Fang Kang, Haoyu Chen}
\address{
   Center for Machine Vision and Signal Analysis, University of Oulu, Finland}
\email{\{sofoklis.kakouros,fang.kang,chen.haoyu\}@oulu.fi}

\begin{document}

\maketitle
\begin{abstract}
This work presents iMiGUE-Speech, an extension of the iMiGUE dataset that provides a spontaneous affective corpus for studying emotional and affective states. The new release focuses on speech and enriches the original dataset with additional metadata, including speech transcripts, speaker-role separation between interviewer and interviewee, and word-level forced alignments. Unlike existing emotional speech datasets that rely on acted or laboratory-elicited emotions, iMiGUE-Speech captures spontaneous affect arising naturally from real match outcomes. To demonstrate the utility of the dataset and establish initial benchmarks, we introduce two evaluation tasks for comparative assessment: speech emotion recognition and transcript-based sentiment analysis. These tasks leverage state-of-the-art pre-trained representations to assess the dataset’s ability to capture spontaneous affective states from both acoustic and linguistic modalities. iMiGUE-Speech can also be synchronously paired with micro-gesture annotations from the original iMiGUE dataset, forming a uniquely multimodal resource for studying speech–gesture affective dynamics. The extended dataset is available at \url{https://github.com/CV-AC/imigue-speech}.



\end{abstract}

\vspace{-2mm}
\section{Introduction}
Prosody—how we modulate pitch, loudness, rhythm, and timing in speech—is one of the most powerful carriers of affect, yet much of our current understanding and evaluation of prosodic emotion cues still relies on acted, scripted, or otherwise controlled data that poorly reflect how people actually speak in emotionally charged situations. 

A dataset that captures speech in genuinely affective, naturalistic contexts—such as athletes reacting to wins and losses in post-match interviews—provides a rare opportunity to study how emotions and sentiments are expressed when they truly matter to the speakers \cite{kakouros25b_interspeech}. In such settings, prosodic patterns emerge spontaneously rather than being performed to match a label set, revealing subtle shifts in pitch, intensity, speech rate, and pausing that may not appear in laboratory recordings. This makes the dataset crucial not only for building and evaluating robust models of prosody-based emotion and sentiment analysis, but also for advancing our theoretical understanding of how affect is encoded in real-world speech, where it is shaped by several factors such as social norms, stress, fatigue, and context.

Traditionally, research on the prosody of affective speech has been extensive, with many studies showing that prosodic characteristics such as fundamental frequency (F0) are influenced by affective state and, more generally, that emotions modulate prosody \cite{larrouy2025sound,banse1996acoustic,scharenborg18b_speechprosody}. Findings are diverse, suggesting that speakers use clusters of acoustic-prosodic cues to encode emotions \cite{brave2007emotion,picard2000affective}. For example, happiness/joy is often associated with higher F0, a wider pitch range, livelier intonation, and a faster speech rate, frequently accompanied by increased intensity relative to neutral speech \cite{juslin2003communication,scherer2003vocal}.

Overall, affective influence can be described in terms of (a) emotion-specific acoustic encoding and (b) dimensional effects (valence/arousal) (see, e.g., \cite{scherer2003vocal}). The former refers to the use of combinations of acoustic-prosodic features to signal perceived discrete emotional categories such as happiness, anger, and sadness. The latter conceptualizes emotions not as named categories (e.g., anger, sadness, joy) but as locations on a small number of continuous scales. The two most common are valence (how pleasant or unpleasant the emotion feels) and arousal (how activated or energized the emotion is). Both dimensions have acoustic-prosodic correlates, with arousal showing a particularly robust overall impact \cite{scherer2003vocal,belyk2014acoustic}.

In this work, we introduce a speech dataset featuring prosodic variations that are elicited not in a controlled experiment but in the natural setting of post-match interviews. The athletes speak in a highly charged context, with sentiment labels determined by match outcomes. To demonstrate the utility of the dataset and establish initial performance benchmarks for the iMiGUE-Speech extensions, we operationalize the application scenarios outlined in Section 3 into two concrete evaluation tasks: Speech Emotion Recognition (SER) \cite{kakouros2023speech,kakouros25_interspeech,stafylakis2023extracting} and transcript-based sentiment analysis \cite{wankhade2022survey}. These tasks leverage state-of-the-art pre-trained representations to assess the dataset’s capacity to capture spontaneous affective states from both acoustic and linguistic modalities.

\vspace{-2mm}
\section{Materials and data processing}

\subsection{iMiGUE dataset}
The iMiGUE (Identity-free Micro-Gesture Understanding and Emotion Analysis) dataset is a specialized video dataset designed for studying micro-gestures \cite{chen2023smg,chen2019analyze}
and their relationship to emotional states while ensuring privacy \cite{liu2021imigue}. Unlike conventional emotion recognition datasets that focus on facial expressions or speech, iMiGUE captures subtle, unintentional body movements that reflect internal emotions. The dataset consists of 359 videos from Grand Slam tennis post-match press conferences, featuring 72 players (36 male, 36 female) from 28 countries. Each video is annotated with micro-gesture categories at the clip level and emotional states (positive or negative) at the video level, based on match outcomes. Importantly, it is identity-free: while faces are masked or removed to protect personal identities, voices are preserved.

While iMiGUE provides a rich set of 18,499 annotated micro-gesture samples across 32 distinct categories, this work expands the dataset to extract audio and metadata from the speech signals.

\vspace{-2mm}
\subsection{Data processing}
\begin{figure}[t]
    \centering
    \includegraphics[width=0.5\columnwidth]{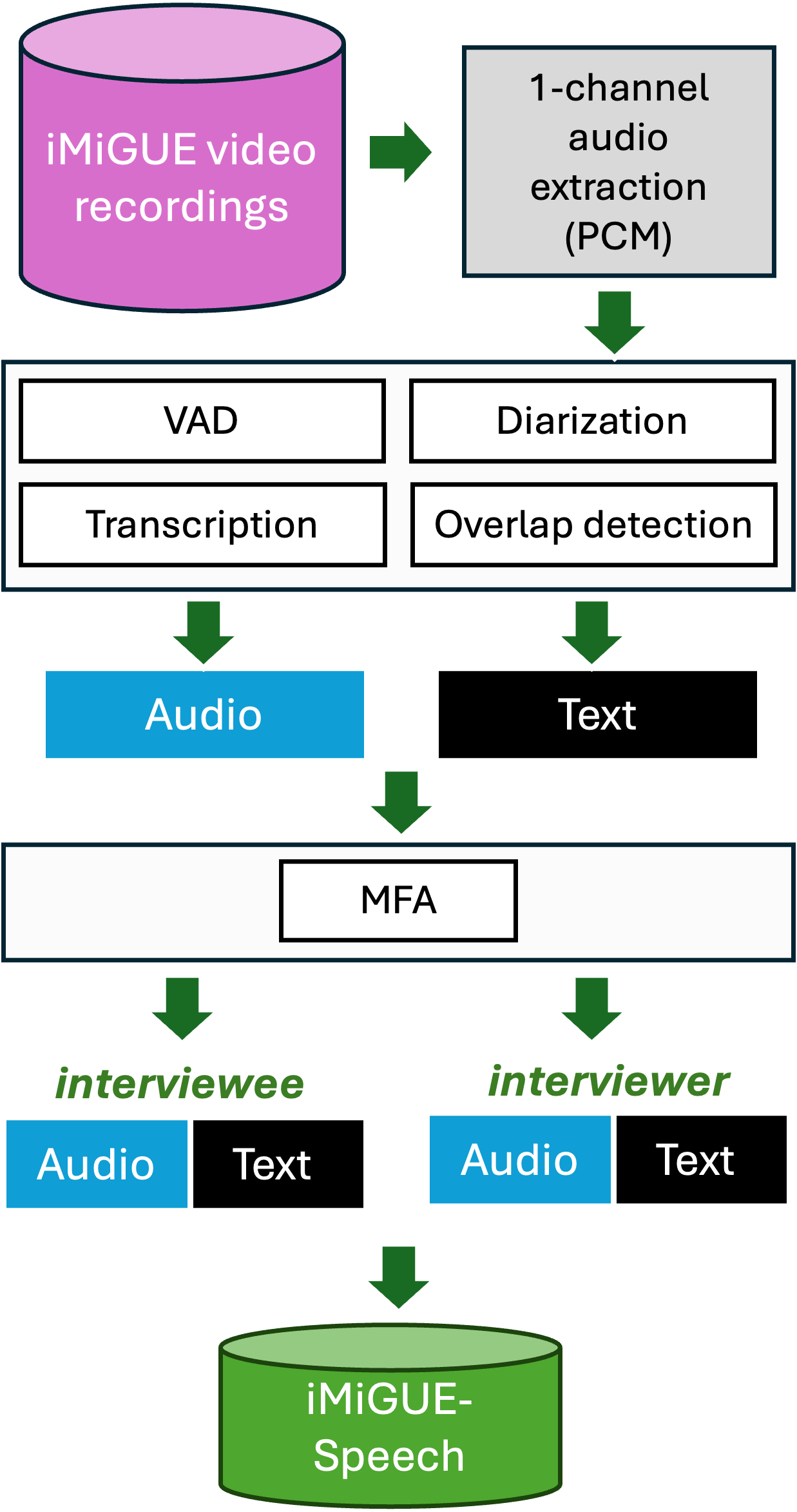}
    \caption{iMiGUE-Speech processing pipeline.}
    \label{fig:imigue_speech_overview}
\vspace{-6mm}
\end{figure}

The main challenge in working with the iMiGUE dataset is to automatically extract the audio segments in which the athlete is actively responding to questions or speaking in general. Although the iMiGUE dataset provides manually annotated gesture labels for the athletes, it does not include speaker-turn annotations or explicit information about who is speaking at each time instant. As a result, the raw recordings typically contain a mixture of interviewer and interviewee speech, overlapping speech, background noise, and silence. To handle this, we design an automatic audio-processing pipeline to process the original video to the final speaker-specific segments.

First, the audio track is extracted from each video recording and converted to a consistent format (e.g., single-channel, fixed sampling rate). We then apply a suite of speech-processing models using the \texttt{pyannote.audio} toolkit: speaker diarization, overlap detection, and voice activity detection (VAD). Speaker diarization partitions the recording into homogeneous segments, each associated with an anonymous speaker label (e.g., \texttt{SPEAKER\_00}, \texttt{SPEAKER\_01}). Overlap detection identifies time intervals where multiple speakers talk simultaneously, which is important for correctly attributing speech and for later filtering. VAD is used to remove non-speech regions, such as silence or background noise, and to focus subsequent processing only on segments that contain speech.

Transcriptions are generated using the Whisper Large model, with the language flag explicitly set to English to ensure consistent English transcripts even in the presence of accents or background noise. For each recording, the outputs of these steps are consolidated into a single TextGrid file in the Praat-compatible format. Each type of annotation is stored in its own tier: one tier for speaker labels (diarization), one for overlapping speech segments, one for VAD-based speech regions, and one for the textual transcriptions aligned at the segment level. This multi-tier structure allows all annotations to be viewed and edited jointly in a single time-aligned representation.

Additionally, forced alignment is performed using the Montreal Forced Aligner (MFA) to obtain precise word-level boundaries \cite{mcauliffe2017montreal}. MFA takes as input the raw audio and the Whisper-generated transcripts and produces time-stamped alignments for individual words (and optionally phones). These word-level boundaries are stored in a separate TextGrid file, again in tiered form, which can be directly inspected or used as input for downstream modeling. The resulting alignments enable more fine-grained analyses, such as synchronizing gestures with individual words or syllables.

To distinguish the athlete from the journalists, we compute the total speaking time for each diarized speaker label across the entire recording. We assume that the athlete speaks for the longest cumulative duration, which is consistent with the interview setting where the interviewee provides extended answers while journalists ask shorter questions. Accordingly, the speaker with the longest speech duration is identified as the athlete. Using this mapping, all segments attributed to the athlete are extracted from the full audio and saved as separate audio clips. Each extracted athlete segment is indexed sequentially (e.g., \texttt{segment\_001}, \texttt{segment\_002}, \dots) according to its temporal order in the original recording.

While the original interview recordings can last several minutes, the extracted segments are typically only a few seconds long, corresponding to individual answers or parts of longer answers. For completeness, we also retain the non-athlete speech segments, which correspond primarily to the journalists. In the final step, all segments are organized into two disjoint sets: one for the athlete (interviewee) and one for the journalists (interviewers). This organization ensures a clear separation between the two roles and provides a clean, speaker-specific set of audio segments that can be directly paired with gesture annotations for subsequent analysis and modeling.

\vspace{-2mm}
\subsection{Extracted metadata}

The pipeline (Figure \ref{fig:imigue_speech_overview}) adds several layers of speaker- and speech-related metadata to iMiGUE to compensate for the lack of speaker-turn annotations in the original release: it extracts and normalizes audio, then uses \texttt{pyannote.audio} to generate speaker diarization labels (anonymous speakers), overlap segments, and VAD-based speech regions, which are consolidated into a multi-tier Praat TextGrid alongside segment-level English transcriptions produced by Whisper Large (see also Table \ref{tab:imigue_audio_metadata_detailed_twocol}). It further augments the dataset with a separate word-level alignment TextGrid created by the Montreal Forced Aligner using the raw audio and Whisper transcripts. Finally, it derives role metadata by identifying the athlete as the diarized speaker with the longest cumulative speaking time, mapping anonymous speaker IDs to roles (athlete vs. journalists), and exporting speaker-specific audio clips indexed in temporal order (e.g., \texttt{segment\_001}, \texttt{segment\_002}), yielding two disjoint, clean sets of segments that can be paired with gesture labels for downstream analysis.


\begin{table}[t]
\centering
\scriptsize
\setlength{\tabcolsep}{3pt}
\renewcommand{\arraystretch}{0.95}
\caption{Audio metadata/annotations for iMiGUE-Speech.}
\label{tab:imigue_audio_metadata_detailed_twocol}
\begin{tabularx}{\columnwidth}{@{}l l >{\raggedright\arraybackslash}X@{}}
\toprule
\textbf{Type} & \textbf{Tool} & \textbf{Added metadata / output} \\
\midrule
Audio standardization & ffmpeg &
Extracted audio; normalized format (1-channel PCM, fixed sampling rate). \\
Speaker diarization & \texttt{pyannote.audio} &
Speaker-labeled time segments (e.g., \texttt{SPEAKER\_00}). \\
Overlap detection & \texttt{pyannote.audio} &
Intervals of simultaneous speakers. \\
VAD & \texttt{pyannote.audio} &
Speech regions for removing silence/background noise. \\
Segment-level ASR & Whisper Large &
English transcripts aligned to speech segments. \\
Segment-level TextGrid & Praat format &
Unified tiers: diarization, overlap, VAD, transcripts. \\
Word-level alignment & MFA &
Word boundaries from audio + Whisper transcripts. \\
Word-level TextGrid & MFA output &
Separate tiered TextGrid with word and phone alignments. \\
Role identification & Heuristic &
Longest cumulative speaking time mapped to athlete. \\
Speaker-specific clips & Custom &
Disjoint athlete vs.\ journalist audio segment sets. \\
Segment indexing & Custom &
Sequential IDs (e.g., \texttt{segment\_001}). \\
\bottomrule
\end{tabularx}
\vspace{-4mm}
\end{table}

\vspace{-2mm}
\section{Corpus Applications}


\subsection{Speech-based affective and paralinguistic analysis}

The athlete-only audio segments, with precise time boundaries and word-level alignment, support a range of speech-focused affective and paralinguistic studies. For example, models can infer whether the athlete has won or lost based purely on vocal characteristics as a proxy for positive vs. negative emotional state. Beyond this, the corpus enables estimation of continuous arousal or activation, as well as more specific stress or tension levels across different parts of the interview. Fine-grained tasks can explore how emotion shifts from answer to answer and how global affective descriptors such as overall emotion, arousal, and valence evolve over the full interview.

\vspace{-2mm}
\subsection{Dialogue- and interaction-based analysis}

Thanks to diarization and speaker-role separation, the corpus can be used to study the structure and dynamics of interview dialogues. It allows automatic identification of whether a given segment belongs to the athlete or to a journalist, and whether a segment functions as a question or an answer. Turn-taking patterns—who speaks next, and after what kind of pause—can be analyzed to characterize conversational flow in high-pressure interview settings. The detailed time alignments also make it possible to investigate spontaneous phenomena such as filled pauses and disfluencies, and to evaluate automatic speech recognition specifically in this domain, comparing performance across speakers and interaction contexts.

\vspace{-2mm}
\subsection{Text-based affective and semantic analysis}

Using the provided transcripts, the corpus also supports text-centric applications. These include sentiment prediction at the level of individual answers or entire interviews, capturing how the athlete’s wording reflects positive, negative, or mixed emotional states. More broadly, the transcripts allow analyses and modeling of how athletes narrate their experiences, how emotional content is conveyed through lexical choices, and how these textual signals relate to the vocal and gestural modalities. 

\vspace{-2mm}
\section{Evaluated Tasks}
To empirically demonstrate the utility of the iMiGUE-Speech extensions and to establish initial performance benchmarks, we operationalize the application scenarios outlined in Section 3 into two concrete evaluation tasks: Speech Emotion Recognition (SER) and Transcript-based Sentiment Analysis. These tasks leverage state-of-the-art pre-trained representations to assess the dataset's capacity for capturing spontaneous affective states from both acoustic and linguistic modalities. All evaluations reported in this work are conducted on the segmented interviewee (athlete) audio tracks and their corresponding transcripts.

\vspace{-2mm}
\subsection{Speech Emotion Recognition}
Consistent with the analysis potential discussed in Section 3.1, this task focuses on predicting the athlete's emotional state solely from the acoustic signal. Given the complex nature of spontaneous post-match interviews, we evaluate SER performance through two complementary paradigms: dimensional regression and categorical classification.

\subsubsection{Dimensional Emotion Regression}

Spontaneous speech is characterized by subtle fluctuations in affect that are often better captured by continuous dimensions rather than discrete labels. We establish a benchmark for predicting three core affective dimensions:
\begin{itemize}
    \item \textbf{Arousal}: Reflecting the level of physiological activation or intensity (e.g., calm vs.\ excited).
    \item \textbf{Dominance}: Reflecting the perceived sense of control or power in the emotional state (e.g., submissive vs.\ confident).
    \item \textbf{Valence}: Reflecting the qualitative pleasantness of the state (e.g., negative vs.\ positive).
\end{itemize}

\textbf{Benchmark Implementation.}
We utilize a robust \textit{Wav2Vec~2.0} architecture~\cite{baevski2020wav2vec} as the acoustic backbone. Specifically, we employ a large-scale model pre-trained on broadly sourced speech data and subsequently fine-tuned on the MSP-Podcast corpus for dimensional emotion recognition. This approach leverages the model's ability to extract context-rich representations that are robust to the acoustic variability found in press conference settings.

\subsubsection{Categorical Emotion Classification}

To provide a comparison with traditional SER literature, we also define a classification benchmark. This task involves mapping the speech segments into discrete emotional categories, which can serve as high-level descriptors of the athlete's reaction to the match outcome.

\textbf{Benchmark Implementation.} We employ the \textit{WavLM} framework~\cite{Goncalves_2024}, which represents the state of the art in self-supervised speech processing. We utilize a model variant fine-tuned specifically for categorical emotion recognition tasks (consistent with recent SER challenges such as the Odyssey benchmark). This baseline tests the transferability of large-scale speech representations to the specific domain of sports interviews.

\subsection{Transcript-based Sentiment Analysis}

Complementing the acoustic analysis, this task targets the linguistic content of the interaction as described in Section~3.3. It aims to quantify the sentiment polarity embedded in the athlete's verbal responses.
The objective is to classify each transcript segment into \textbf{Positive} or \textbf{Negative} sentiment polarity. This binary classification serves as a proxy for the athlete's explicit evaluation of their performance and the match result.

\textbf{Benchmark Implementation.}
We utilize a \textit{RoBERTa-large} language model~\cite{hartmann2023}, fine-tuned on a diverse aggregation of English sentiment datasets (including reviews and social media text). By using a model trained on varied text sources, we aim to leverage robust semantic representations that can generalize to the domain vocabulary and informal grammatical structures often present in spontaneous spoken transcripts.

\vspace{-2mm}
\section{Results and Discussion}
This section presents the affective patterns observed across athlete interviews after aggregating all segment-level predictions into a single profile for each session ID. Such ID-level representations enable a direct comparison between winning and losing athletes, revealing how match outcome shapes their overall emotional expression. Three complementary affective signals are examined: continuous affective dimensions, distributions of discrete emotion categories, and sentiment polarity extracted from the textual transcripts. Together, these results provide a coherent view of outcome-related affective tendencies in spontaneous post-match speech.

\subsection{Dimensional Affective Estimates}
\begin{table}[t]
\centering
\caption{Dimensional affective statistics for winning and losing athletes.
“Inter Level’’ reflects the across-interview average of each dimension, while “Intra Variability’’ reflects the average within-session fluctuation (larger values in each column are bolded).}
\label{tab:dimensional_results}
\resizebox{\linewidth}{!}{
\begin{tabular}{lcccccc}
\toprule
 & \multicolumn{2}{c}{\textbf{Arousal}} 
 & \multicolumn{2}{c}{\textbf{Dominance}} 
 & \multicolumn{2}{c}{\textbf{Valence}} \\
\cmidrule(r){2-3}\cmidrule(r){4-5}\cmidrule(r){6-7}
\textbf{Group} & \shortstack{Inter \\ Level} & \shortstack{Intra \\  Variability}
               & \shortstack{Inter \\ Level} & \shortstack{Intra \\  Variability}
               & \shortstack{Inter \\ Level} & \shortstack{Intra \\  Variability} \\
\midrule
Lose & 0.3468 & \textbf{0.0787} & 0.4139 & \textbf{0.0696} & 0.5025 & 0.1166 \\
Win  & \textbf{0.3690} & 0.0754 & \textbf{0.4342} & 0.0655 & \textbf{0.5361} & \textbf{0.1202} \\
\bottomrule
\end{tabular}
}
\vspace{-2mm}
\end{table}

Dimensional affective estimates offer a structured view of how emotional tendencies differ between winning and losing athletes. Each interview is characterized using two complementary measures: Inter Level, which reflects the overall affective tendency of a session, and Intra Variability, which captures the extent to which emotional expression fluctuates across its constituent segments.

Inter Level is obtained by averaging the segment-level predictions of each session ID and then computing the mean of these ID-level values within the Win and Lose groups. This measure represents the typical affective state expressed by athletes over the course of their full interview. Intra Variability, by contrast, is derived by computing the standard deviation of segment-level predictions within each session and averaging these values across groups, providing insight into the stability or dynamism of emotional expression.

Table~\ref{tab:dimensional_results} summarizes the resulting group-level patterns. Winners show higher valence and dominance levels, indicating a more positive and confident overall affective state, while arousal levels remain comparatively close. Losing athletes display slightly larger valence variability, suggesting more dynamic internal affective shifts within their interviews. Together, these measures highlight differences not only in overall affective levels but also in the temporal stability of emotional expression following different match outcomes.

\subsection{Discrete Emotion Category Distributions}
Categorical emotion predictions provide an additional perspective on the affective differences between winning and losing athletes. By identifying the dominant emotion category for each interview, we obtain a session-level label that reflects the primary emotional tone expressed during the interaction. The distribution of these dominant categories across the two outcome groups is presented in Table~\ref{tab:cat_emotion_results}.

\begin{table}[t]
\centering
\caption{Proportion (\%) of dominant categorical emotions for winning and losing athletes (larger values in each column are bolded).}
\label{tab:cat_emotion_results}
\resizebox{\linewidth}{!}{
\begin{tabular}{lcccccccc}
\toprule
\textbf{Group} & 
\textbf{Angry} & 
\textbf{Sad} & 
\textbf{Happy} & 
\textbf{Surprise} & 
\textbf{Fear} & 
\textbf{Disgust} &
\textbf{Contempt} &
\textbf{Neutral} \\
\midrule
Lose & 0.0 & \textbf{85.2} & 5.9 & 0.0 & \textbf{2.0} & 0.0 & 0.0 & 6.9 \\
Win  & 0.0 & 64.7 & \textbf{15.5} & 0.0 & 0.8 & 0.0 & 0.0 & \textbf{19.0} \\
\bottomrule
\end{tabular}
}
\vspace{-4mm}
\end{table}

Clear distinctions emerge between winners and losers. Losing athletes show a pronounced concentration in the Sad category (85.2\%), accompanied only by small proportions of Happy, Fear, and Neutral, resulting in a strongly negative affective profile. The winning group, in contrast, exhibits a more balanced emotional distribution: although Sad remains the most frequent category (64.7\%), its dominance is substantially reduced, and both Happy (15.5\%) and Neutral (19.0\%) categories occur more often, indicating a shift toward more positive or emotionally neutral states.

Emotion categories such as Angry, Surprise, Disgust, and Contempt do not appear as dominant emotions in either group. This absence aligns with the communicative norms of post-match press interviews, where athletes typically maintain composure and rarely express overtly hostile or highly animated emotions.

Overall, the categorical results reinforce the outcome-related affective patterns observed in the dimensional estimates: losing athletes tend to express more negative emotions, while winning athletes present a broader and more positive emotional spectrum, consistent with their respective match outcomes.

\vspace{-2mm}
\subsection{Transcript-Based Sentiment Polarity}
\begin{table}[t]
\centering
\caption{Proportion (\%) of session-level sentiment polarity across winning and losing athletes (larger values are bolded).}
\label{tab:text_sentiment_results}
\resizebox{0.7\linewidth}{!}{
\begin{tabular}{lcc}
\toprule
\textbf{Group} & \textbf{Negative (\%)} & \textbf{Positive (\%)} \\
\midrule
Lose & 12.9 & \textbf{87.1} \\
Win  & 4.6  & \textbf{95.4} \\
\bottomrule
\end{tabular}
}
\vspace{-6mm}
\end{table}

Text-based sentiment analysis offers an additional perspective on how athletes verbally frame their post-match reflections. Each interview transcript is assigned a session-level polarity label (Positive or Negative) based on the average sentiment score across all textual segments. The distribution of these labels within the Win and Lose groups is shown in Table~\ref{tab:text_sentiment_results}.

The patterns are consistent with the affective trends observed in the speech-based analyses. Losing athletes exhibit a higher proportion of Negative sentiment (12.9\%) relative to winning athletes (4.6\%), reflecting more self-critical or dissatisfied verbal expressions following a loss. Conversely, the majority of interviews in both groups are labeled Positive, but the dominance of Positive sentiment is substantially stronger among winners (95.4\% vs. 87.1\%). This indicates that winning interviews tend to contain more upbeat, confident, or forward-looking language.

Overall, the sentiment results align closely with both the dimensional and categorical emotion findings: losing athletes express more negative or dissatisfied attitudes, while winning athletes adopt a more positive evaluative tone, reinforcing the systematic affective differences associated with match outcomes.

\vspace{-2mm}
\section{Conclusions}

This work presented iMiGUE-Speech, an extension to iMiGUE providing a richer resource for studying spontaneous affect by pairing speech with detailed transcripts, role separation, and word-level alignments. The baseline SER and transcript-based sentiment tasks show that the dataset can capture meaningful outcome-related affective patterns in post-match interviews. The dataset will be publicly available to support future research in affect and related fields.


\section{Acknowledgements}

This work was supported by the University of Oulu (InfoTech) and the Research Council of Finland (PROFI7 352788, Research Fellow 371019 projects). Computational resources were provided by CSC – IT Center for Science, Finland.

\bibliographystyle{IEEEtran}

\bibliography{mybib}


\end{document}